# Reproducibility in Recommender Systems: A Survey

ALAN SAID, University of Gothenburg, Sweden
ALEJANDRO BELLOGÍN, Universidad Autónoma de Madrid, Spain

Reproducibility has become a cornerstone of credible recommender systems research, driven by growing concerns about the reliability and generalizability of experimental results. In response, the ACM RecSys conference introduced a dedicated Reproducibility Track in 2020 to encourage rigorous, transparent, and repeatable research. This paper presents a structured analysis of the track from 2020 to 2025, covering 51 accepted papers. We classify contributions by type and analyze common patterns in datasets, algorithms, frameworks, and evaluation practices, with the goal of understanding how reproducibility is operationalized in practice within the community.

Our findings reveal three main trends. First, the track has expanded in scope, evolving from a focus on reproduction and replication to include benchmarking, resources, and methodological contributions. Second, reproducibility papers exhibit a consistent methodological profile, relying on a limited set of datasets, algorithms, and evaluation protocols. Third, reproducibility in practice often involves extending prior experiments rather than strictly replicating them, with studies frequently introducing additional models or evaluation criteria. Overall, reproducibility work has improved transparency and documentation, but has had limited impact on methodological diversity, highlighting a gap between the conceptual definition of reproducibility and its implementation.

## 1 Introduction

Reproducibility has become a central concern in recommender systems research, reflecting broader discussions about the reliability, transparency, and cumulative nature of empirical science. As recommender systems have grown in complexity, incorporating deep learning models, large-scale datasets, and increasingly intricate evaluation pipelines, the difficulty of reproducing published results has increased accordingly. In response, the ACM Conference on Recommender Systems (RecSys) introduced a dedicated Reproducibility Track in 2020, with the goal of fostering transparent, rigorous, and repeatable research practices.

Early work in the field has repeatedly highlighted shortcomings in experimental methodology and reporting [9, 34, 60]. Studies have shown that weak baselines, insufficient hyperparameter tuning, and inconsistent evaluation protocols can lead to misleading conclusions about progress [17, 59, 64]. Concrete examples include, e.g., Beel et al. [8] which identified several *determinants* that affect the reproducibility of recommender systems experiments, including user characteristics, datasets, weighting schemes, and time. These concerns have motivated not only reproducibility efforts

Authors' Contact Information: Alan Said, alan@gu.se, University of Gothenburg, Gothenburg, Sweden; Alejandro Bellogín, alejandro.bellogin@uam.es, Universidad Autónoma de Madrid, Madrid, Spain.

such as the guidelines for accountable and reproducible research by Bellogín and Said [10], but also a broader line of work focusing on recommender systems evaluation. For example, Bauer et al. [7] provide a systematic review of evaluation-focused research, showing that the field relies heavily on a narrow set of datasets (e.g., MovieLens, Amazon) and metrics (e.g., nDCG, Precision, Recall), with limited attention to methodological diversity or beyond-accuracy evaluation. While this work does not cover the full spectrum of recommender systems research, it offers a useful reference point for understanding common methodological patterns in the field.

Reproducibility and evaluation research thus represent two distinct but related perspectives on the methodological foundations of recommender systems. Evaluation research focuses on how systems *should* be assessed, often proposing new metrics, protocols, or experimental designs. In contrast, reproducibility research examines how empirical findings *hold up in practice*, by re-running, re-implementing, or extending prior work under transparent and controlled conditions. Understanding how these perspectives relate is important for assessing the robustness of current research practices and identifying opportunities for methodological improvement.

In this paper, we present a comprehensive analysis of the papers published in the RecSys Reproducibility Track from its inception in 2020 through 2025, covering 51 accepted papers. Beyond a descriptive survey, our goal is to understand how reproducibility research has evolved as a methodological practice within the community, and how it relates to broader trends in recommender systems research. In particular, we analyze the extent to which reproducibility papers reinforce, challenge, or extend commonly used experimental practices.

To this end, we consider and address the following research questions:

**RQ1:** What kinds of contributions has the community produced in the RecSys Reproducibility Track?

**RQ2:** What methodological patterns characterize reproducibility papers in terms of datasets, algorithms, frameworks, and evaluation protocols?

**RQ3:** To what extent do reproducibility papers reflect or diverge from methodological patterns reported in prior studies of recommender systems research?

**RQ4:** To what extent do reproducibility studies replicate, modify, or extend the experimental setups of prior work?

To answer these questions, we conduct a structured review of all papers published in the track between 2020 and 2025 (Section 2), combining quantitative analysis of reported experimental components with a qualitative assessment of contribution types and research practices (Section 3), where the previous questions are addressed. This enables us to systematically characterize how reproducibility is enacted in practice and to identify recurring methodological patterns, quantify reproducibility efforts invested by the community, and suggest implications for the research area (Section 4).

Our findings show that while the Reproducibility Track has significantly improved transparency and artifact availability, it has also undergone a notable shift in scope. In particular, the expansion of accepted contribution types between 2023 and 2025 has broadened the notion of reproducibility to include resources, methodologies, and reflective works. At the same time, reproducibility studies largely rely on the same limited set of datasets, metrics, and evaluation protocols identified in prior work, suggesting that methodological diversity remains constrained.

Overall, this work provides both a retrospective analysis of the RecSys Reproducibility Track and a critical perspective on its role in shaping research practices. By situating reproducibility within the broader landscape of methodological research in recommender systems, we aim to contribute to ongoing discussions on how to achieve more robust, transparent, and comparable empirical results.

## 2 Review Methodology

To investigate the evolution of reproducibility research in the RecSys community, we conducted a structured review of all papers accepted to the RecSys Reproducibility Track between 2020 and 2025 ($n$ = 51). Our process, shown in Fig. 1, follows four stages: scoping, data collection, coding, and analysis.

*Scoping and Selection.* All papers were retrieved from the ACM Digital Library using the official RecSys proceedings indices, ensuring full coverage of the reproducibility track for each year.

*Data Collection.* For each paper, we recorded bibliographic metadata (year, title, authors) and extracted variables relevant to our analysis, including paper length (full or short)[1], contribution type, datasets and algorithms used, evaluation protocols, and artifact availability. Artifact availability was assessed by manually verifying whether source code, datasets, or supplementary materials were provided.

*Coding Procedure.* Each paper was independently reviewed by both authors, applying a predefined coding schema. The schema captures:

(i) contribution type (see Table 2),
(ii) evaluation settings (datasets, metrics, baselines),
(iii) reproduction method (e.g., reuse of original code, reimplementation, or extension to new contexts),
(iv) artifact availability (data, framework, platform), and
(v) use of frameworks or libraries for experimentation.

Discrepancies were resolved through discussion to ensure consistency.

*Analysis.* Based on the coded data, we conducted both quantitative and qualitative analyses. Quantitatively, we examine distributions and trends over time in datasets, algorithms, frameworks, and evaluation practices. Qualitatively, we analyze how reproducibility is operationalized in practice, including how studies reproduce, modify, or extend prior work. This analysis supports RQ1 and RQ2, while also enabling comparisons to prior methodological studies (RQ3) and an assessment of reproducibility efforts (RQ4).

### 2.1 Reproducibility Track Overview

Since its inception in 2020, the RecSys Reproducibility Track[2] has undergone notable changes in scope and structure (Table 1). Initially, the track focused narrowly on *replicability* and *reproducibility* studies—clearly distinguishing between repeating experiments with original code and data (replicability) versus conducting experiments in new contexts or with extended setups (reproducibility). Authors were required to submit detailed artifacts, and the review process emphasized the clarity, availability, transparency, and completeness of code and data. The track was reviewed single-blind, and contributions were strictly required to originate from researchers other than the original authors of the reproduced work.

By 2022, the ACM terminology standardization initiative[3] led to a reversal of the definitions of "reproducibility" and "replicability" in the CFP. That year also introduced a *rebuttal phase*, allowing authors to clarify misunderstandings in reviews focused on the reproducibility artifacts.

[1] From 2020–2023, RecSys did not distinguish paper types by page limits. We assigned "short" (up to 6 pages in ACM double column format) and "full" labels based on length.
[2] Links to track CFP's 2020, 2021, 2022, 2023, 2024, 2025, 2026
[3] https://www.acm.org/publications/badging-terms

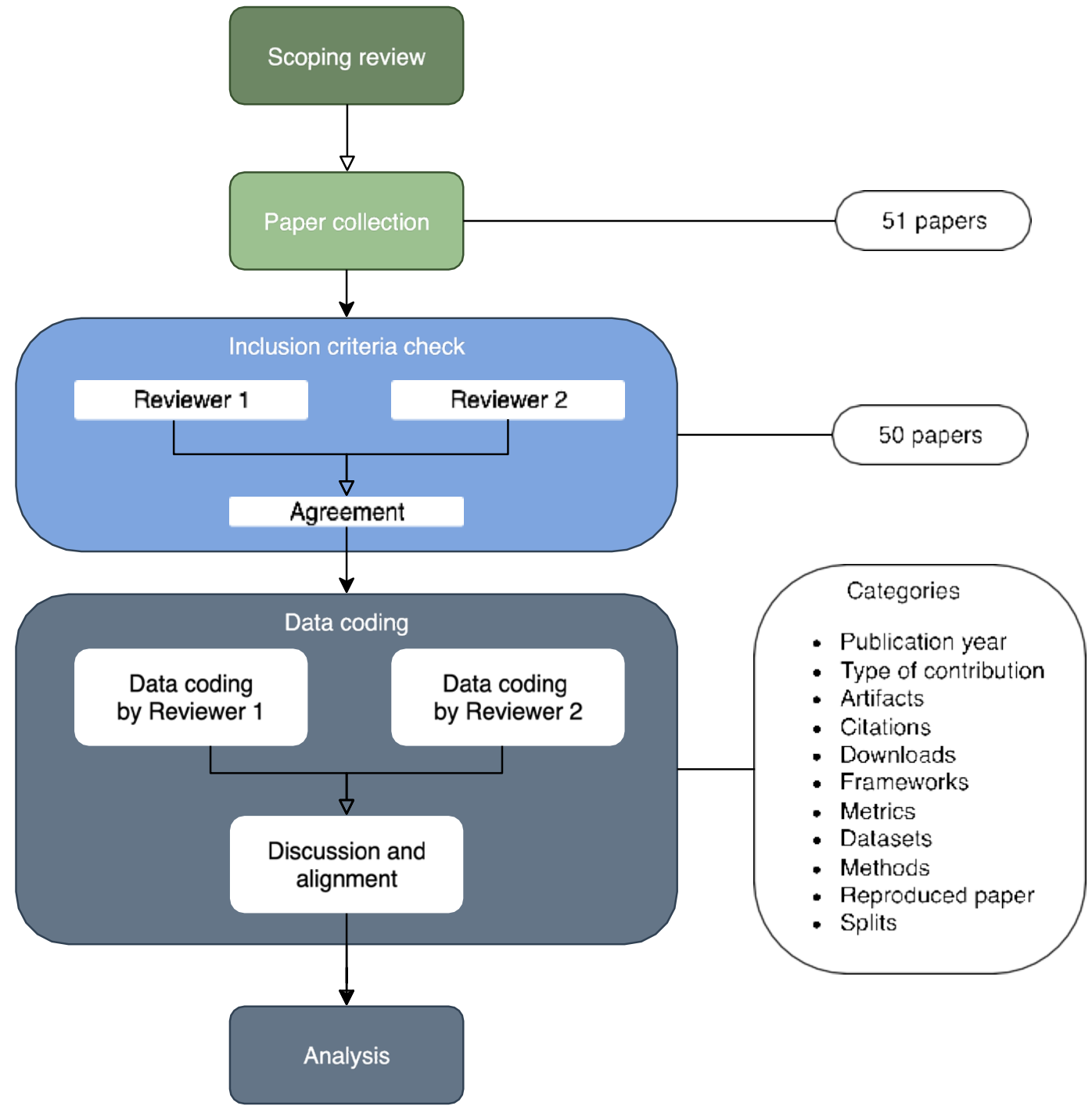

Fig. 1. Review procedure of the papers.

Over time, the scope of the track broadened. In 2023, new contribution types were introduced, including *resource papers* (e.g., datasets or frameworks), *methodology papers* (e.g., novel evaluation protocols), and *reflective contributions* (e.g., surveys and guidelines). This expansion recognized that enabling reproducibility extends beyond repeating experiments to include enabling infrastructure and methodological reflection.

This broader scope was formalized in 2024 and maintained in 2025. The reviewing process also shifted to *double-blind* for all but resource papers. The CFPs also emphasized the use of community-curated frameworks and datasets (available on RecSys GitHub) and provided detailed guidelines for artifact preparation and anonymization.

In 2026, the track was revised again, narrowing its scope to reproducibility/replicability and resource papers, and re-emphasizing the requirement that reproducibility studies provide new empirical insights beyond re-running prior experiments. This realignment reflects an effort to restore the track's original focus while retaining lessons learned from its expansion.

### 2.2 Coding and Categorization

For each paper, we collected metadata including publication year, citation count, number of downloads (as of February 27, 2026), and paper length from the ACM Digital Library (Table 3). In addition, we extracted key elements related to experimental design: the reproduced work (if any), dataset(s) used, evaluation metric(s), recommendation model(s), dataset splitting strategies, contribution type, and presence and nature of associated artifacts.

Table 1. Evolution of the RecSys Reproducibility Track (2020–2026)

| Year | Focus | Key Developments | Papers |
|---|---|---|---|
| 2020 | Replicability & Reproducibility | First edition of the track. Focused on repeating or extending prior work using original code or in new contexts. Single-blind review. Artifacts required. Authors of original works excluded from submission. | 2 |
| 2021 | Consistency | Reinforcement of expectations for materials (code, data). Companion websites encouraged. Review criteria and evaluation rigor emphasized. | 3 |
| 2022 | Terminology Alignment | Switched definitions of "reproducibility" and "replicability" to match ACM-wide usage. Introduced an artifact-centered rebuttal phase. | 3 |
| 2023 | Broadening | Added new paper types (resource, methodology, reflective works), expanding the track beyond strict replication studies. Introduced double-blind review. Repositioned the track as a broader venue for reproducibility-related contributions, including infrastructure, evaluation design, and meta-research. | 7 |
| 2024 | Formalization | Clear classification of paper types. Stronger guidelines for anonymization and artifact hosting. Continued emphasis on community-curated tools and datasets. | 14 |
| 2025 | Consolidation | Maintained the broadened scope introduced in previous years, encompassing reproducibility , resource, methodology, and reflective works. Introduced a dedicated artifact-focused rebuttal phase and reinforced documentation and artifact requirements. | 22 |
| 2026 | Realignment | Revised the track to focus exclusively on reproducibility/replicability and resource papers only. Clarified expectations that reproducibility papers must yield new scientific insights beyond re-running prior experiments. Strengthened artifact requirements, including documentation of hardware configurations and execution environments. Re-emphasized the track's core identity as an empirical reproducibility venue. | |

Table 2. Classification schema for contribution types: benchmarking, evaluation, resource, and survey.

| Type | Description |
|---|---|
| Benchmarking | Papers that compare the performance of algorithms, models, or systems, often by reproducing or extending prior experiments. |
| Evaluation | Papers that focus on validating specific experimental setups or outcomes, typically by repeating key parts of previous studies. |
| Resource | Papers that introduce datasets, libraries, or tools to support future reproducibility efforts. |
| Survey | Papers that summarize the state of reproducibility, present guidelines, or reflect on methodological practices in the field. |

Each paper was assigned a primary contribution type based on its dominant focus. Although some papers exhibit multiple characteristics, we adopt a single-label classification to enable consistent aggregation and analysis. The resulting categorization includes four types: *benchmark*, *evaluation*, *resource*, and *survey*, as summarized in Table 2.

Drawing on this, we can answer **RQ1** *What kinds of works has the community contributed to the RecSys Reproducibility Track?* partly by considering the paper types (Table 2) and the evolution of the reproducibility track over the years (Section 2.1).

To support RQ4, we further annotated papers that explicitly reproduce or replicate prior work, recording the differences between the original and reproduced studies in terms of datasets, algorithms, and metrics. This enables a systematic analysis of how reproducibility efforts modify experimental setups in practice.

All coding was performed independently by both authors. Discrepancies between the initial annotations were subsequently identified and resolved through discussion until consensus was reached, resulting in a single agreed-upon annotation for each item. This consensus-based coding procedure was used to ensure consistency across the dataset and forms the basis for the analyses presented in the following sections.

To support reproducibility of our review, all coded variables and extracted metadata are made available in the supplementary materials.

## 3 Paper Analysis

We now analyze reproducibility practices across the papers published in the RecSys Reproducibility Track between 2020 and 2025. Of the 51 papers published in the track, one was excluded as it did not focus on recommender systems, resulting in a final sample of 50 papers (Table 3).

Our analysis addresses all four research questions. First we characterize the types of contributions in the track and their evolution over time (RQ1). We then examine the methodological components reported in reproducibility papers—frameworks, datasets, algorithms, and evaluation protocols (RQ2). Building on this, we asses how these practices relate to broader methodological patterns reported in prior studies (RQ3). Finally, we analyze how reproducibility studies modify or extend the experimental setups of the original work (RQ4).

The goal of this section is twofold: *(i)* to provide a structured descriptive overview of how reproducibility is operationalized in practice, and *(ii)* to identify recurring patterns, limitations, and opportunities for methodological improvement.

While each subsection focuses on tools and techniques that appear in at least two papers, Table 4 provides a complete overview of the total number of distinct and reused algorithms, datasets, metrics, and frameworks across all papers.

### 3.1 Contribution Types and Track Evolution

Table 5 and Fig. 2 present the distribution of contribution types across the reviewed papers. Benchmarking papers dominate the track, followed by resource papers, while evaluation and survey contributions remain comparatively limited. Fig. 3 presents an overview of the number of citations (average) papers from each year of the track have received, as well as a normalized citation count averaged over the number of years since the papers were published.

Over time, the distribution of contribution types shifts significantly. In the early years (2020–2022), the track is composed primarily of reproduction and replication studies, closely aligned with its original intent. Starting in 2023, however, with the introduction of additional contributions types, such as resource, methodology, and reflective works, leads to a broader and more diverse set of submissions. This expansion is particularly visible in 2024 and 2025, where resource and benchmarking papers account for a substantial share of accepted contributions.

Artifact availability has varied over time, reflecting the evolution of the track's scope. In the first two years (2020–2021), when the call for papers primarily emphasized reproductions and replications, artifacts were less common. Starting in 2022, with the formal inclusion of resource and methodology papers, artifact sharing became more prevalent—particularly in resource papers, where it is now expected. An exception is Sun et al. [69], which, although categorized as a *Survey*, introduced a reproducibility framework.

Submission volume has also grown significantly, with 2023–2025 accounting for the majority of published papers. In 2024 alone, 14 papers were accepted, nearly doubling the cumulative total from the first three years, and this growth continued in 2025 (Fig. 2). This increase coincides with the expansion of the track's scope, suggesting that a broadening of accepted contribution types contributed to higher submission rates.

Table 3. Papers sorted by year, L., C., DL., and Reprod. refer to whether the paper was a short (S) or full (F) paper, the number of citations, the number of downloads (from ACM DL), and the paper being reproduced (if applicable).

| Year | Paper | L. | C. | DL. | Reprod. |
|---|---|---|---|---|---|
| 2020 | Sun et al. [69] | F | 120 | 3,061 | |
| 2020 | Rendle et al. [56] | F | 365 | 17,084 | [25] |
| 2021 | Dallmann et al. [18] | F | 59 | 692 | [37] |
| 2021 | Manzoor and Jannach [45] | S | 11 | 742 | [15] |
| 2021 | Anelli et al. [1] | F | 35 | 1,193 | [56] |
| 2022 | Petrov and Macdonald [51] | F | 50 | 1,931 | [68] |
| 2022 | Rendle et al. [57] | S | 45 | 3,388 | [30] |
| 2022 | Latifi and Jannach [39] | S | 9 | 1,878 | [54] |
| 2023 | Anelli et al. [3] | F | 15 | 1,853 | |
| 2023 | Shehzad and Jannach [64] | S | 22 | 1,648 | |
| 2023 | Bölz et al. [11] | F | 7 | 940 | |
| 2023 | Wang et al. [75] | F | 6 | 605 | |
| 2023 | Lops et al. [42] | F | 5 | 525 | [23] |
| 2023 | Paparella et al. [50] | F | 2 | 1,206 | [67] |
| 2023 | Hidasi and Czapp [28] | F | 14 | 677 | [29] |
| 2024 | Ariza-Casabona et al. [4] | F | 6 | 1,270 | |
| 2024 | Malitesta et al. [44] | F | 4 | 1,418 | |
| 2024 | Gómez et al. [21] | F | 6 | 1,072 | |
| 2024 | Bakhshizadeh et al. [5] | S | 2 | 782 | |
| 2024 | Matrosova et al. [47] | F | 3 | 921 | [40] |
| 2024 | Boratto et al. [13] | F | 5 | 1,382 | [12] |
| 2024 | Vente et al. [73] | F | 31 | 1,802 | |
| 2024 | Heitz et al. [26] | F | 3 | 1,048 | |
| 2024 | Zhang et al. [80] | F | 18 | 2,863 | |
| 2024 | Chen and Lin [16] | F | 1 | 770 | [14] |
| 2024 | Li et al. [41] | F | 4 | 1,151 | |
| 2024 | Irrera et al. [31] | F | 1 | 1,078 | |
| 2024 | Lops et al. [43] | F | 6 | 1,169 | [20] |
| 2024 | Milogradskii et al. [48] | F | 12 | 1,090 | [55] |
| 2025 | Nguyen et al. [49] | F | 0 | 2,589 | |
| 2025 | Zeng et al. [79] | F | 0 | 2,451 | |
| 2025 | Sánchez et al. [70] | F | 0 | 2,174 | |
| 2025 | Martina et al. [46] | F | 0 | 2,620 | |
| 2025 | Pires et al. [52] | F | 0 | 2,265 | |
| 2025 | Kang et al. [33] | F | 0 | 2,548 | |
| 2025 | Vlachou [74] | F | 1 | 2,222 | |
| 2025 | Ballocu et al. [6] | F | 0 | 2,556 | |
| 2025 | Fioretti et al. [19] | F | 0 | 2,831 | [77] |
| 2025 | Ungruh et al. [71] | F | 1 | 2,518 | [72] |
| 2025 | Heitz et al. [27] | F | 0 | 2,502 | |
| 2025 | Zhao et al. [81] | F | 0 | 2,521 | |
| 2025 | Wang et al. [76] | F | 0 | 2,579 | |
| 2025 | Kusano et al. [38] | F | 0 | 3,318 | |
| 2025 | Shehzad and Jannach [65] | S | 0 | 2,456 | |
| 2025 | Spillo et al. [66] | F | 0 | 2,836 | |
| 2025 | Sbandi et al. [62] | F | 0 | 2,435 | |
| 2025 | Isufaj et al. [32] | F | 0 | 2,119 | |
| 2025 | Gusak et al. [22] | F | 2 | 2,728 | |
| 2025 | Shehzad et al. [63] | F | 1 | 2,445 | |
| 2025 | Ploshkin et al. [53] | F | 0 | 2,357 | |

Table 4. Summary of unique and repeated use of algorithms, datasets, metrics, and frameworks across all reviewed papers.

| | Number | Uses |
|---|---|---|
| Algorithms | 186 | 312 |
| Datasets | 81 | 155 |
| Frameworks | 24 | 53 |
| Metrics | 62 | 176 |

Table 5. Categorization of reviewed papers by primary contribution type and the total number of papers of the type. Asterisks (*) denote artifact contributions.

| Type | Papers | nr. |
|---|---|---|
| Benchmarking | [56], [45], [57], [39], [42], [28], [64], [3], [31], [16], [80], [73], [13], [4], [47], [48] [49], [79], [52], [33], [19], [81], [76], [38], [65], [22] | 26 |
| Evaluation | [18], [1], [50], [44], [71] | 5 |
| Resource | [75]*, [11]*, [41]*, [26]*, [5]*, [21]*, [43]*, [70]*, [46]*, [74]*, [6]*, [27]*, [66]*, [62]*, [32]*, [53]* | 16 |
| Survey | [69]*, [51], [63] | 3 |

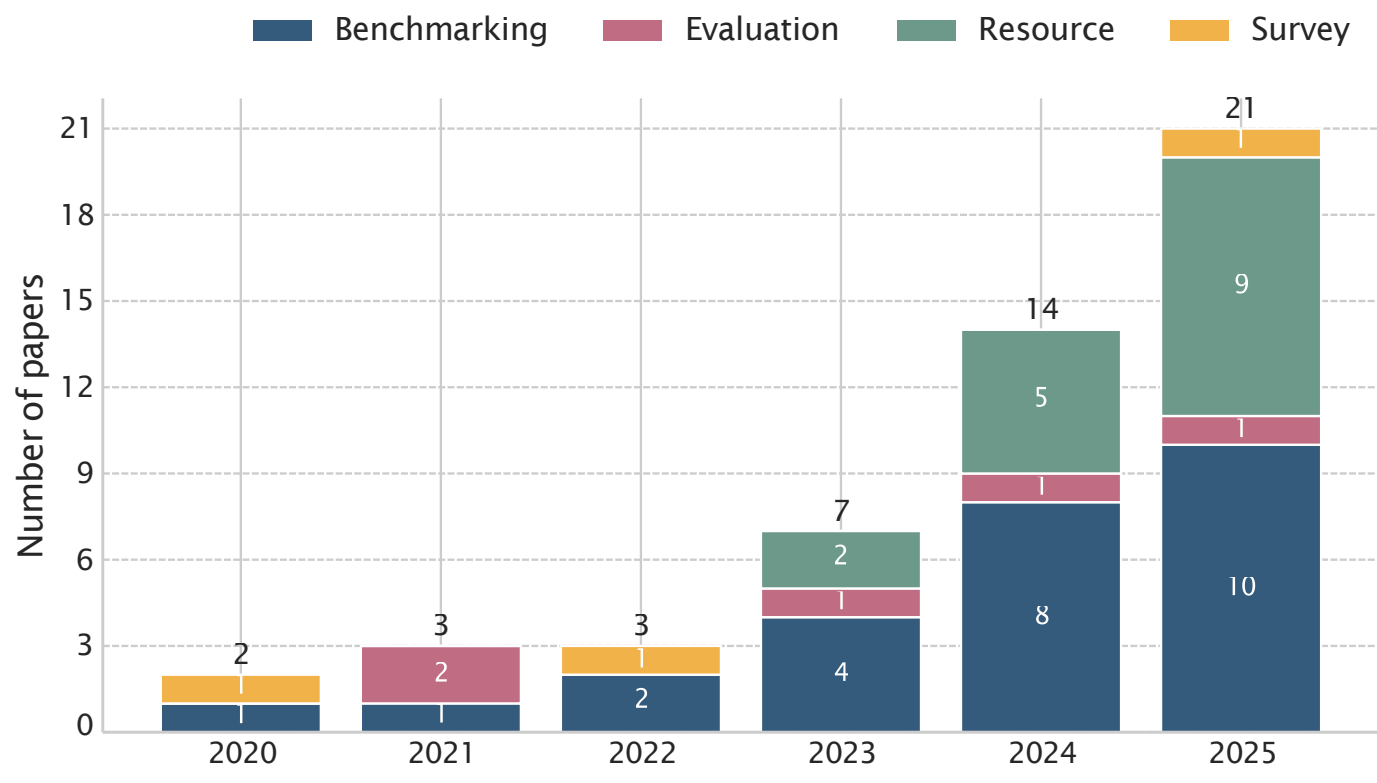

Fig. 2. Distribution of contribution types by year (2020–2025) in the RecSys Reproducibility Track.

Overall, these trends indicate that the Reproducibility Track has evolved from a narrowly defined venue for replication studies into a broader forum for methodological and infrastructural contributions. While this expansion has increased participation and diversity of contributions, it has also shifted the focus away from strict reproduction and replication.

### 3.2 Frameworks

Recommender system frameworks play a central role in enabling reproducibility by standardizing implementations, data handling, and evaluation procedures. Across the reviewed papers, a small number of frameworks dominate, most notably Elliot [2], RecBole [82], and Cornac [61] (Fig. 4).

At the same time, a non-trivial number of papers do not explicitly report the frameworks used. While this may indicate custom implementations, it limits reproducibility and comparability across studies. Even when code is shared, missing information about the underlying framework or experimental environment can introduce ambiguity, as framework-specific defaults and implementation details may significantly affect results.

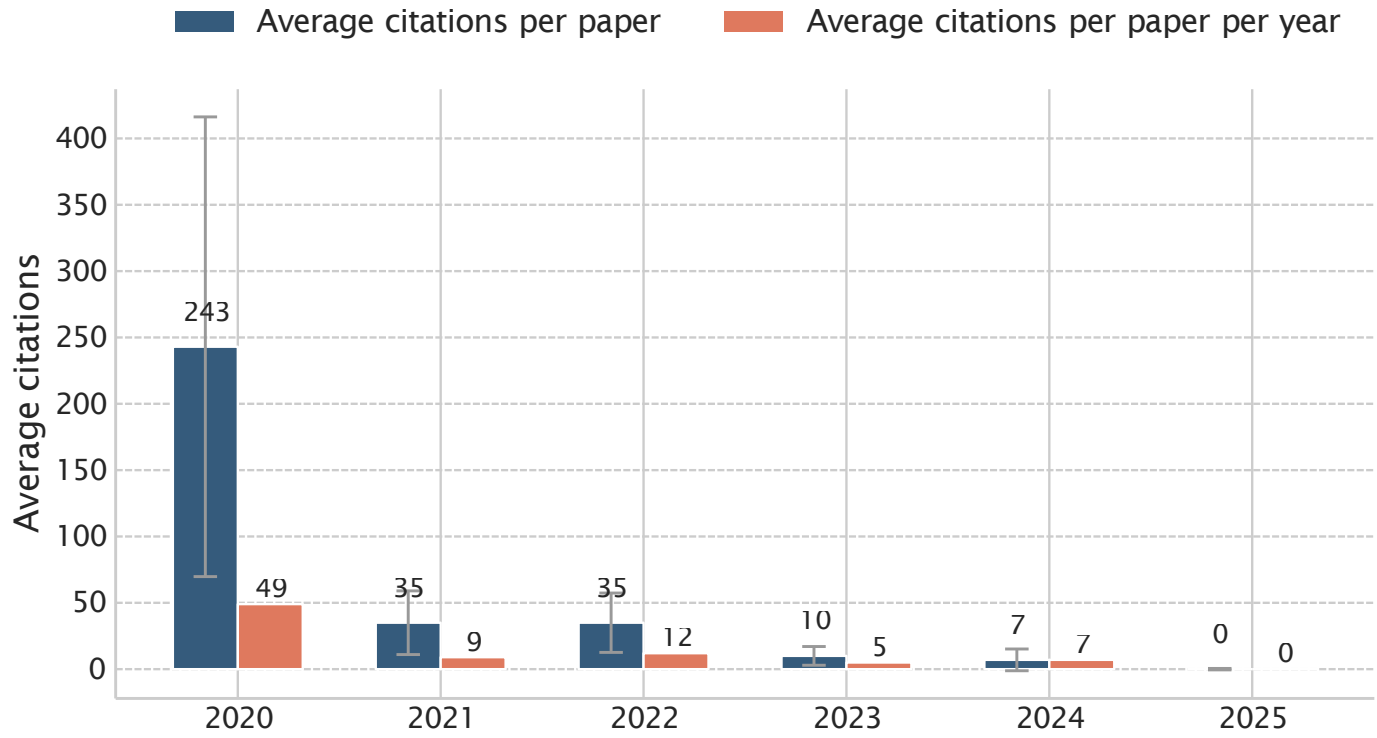

Fig. 3. The average citations per paper (with standard deviation), and a normalized average citations per paper per year since the paper was published.

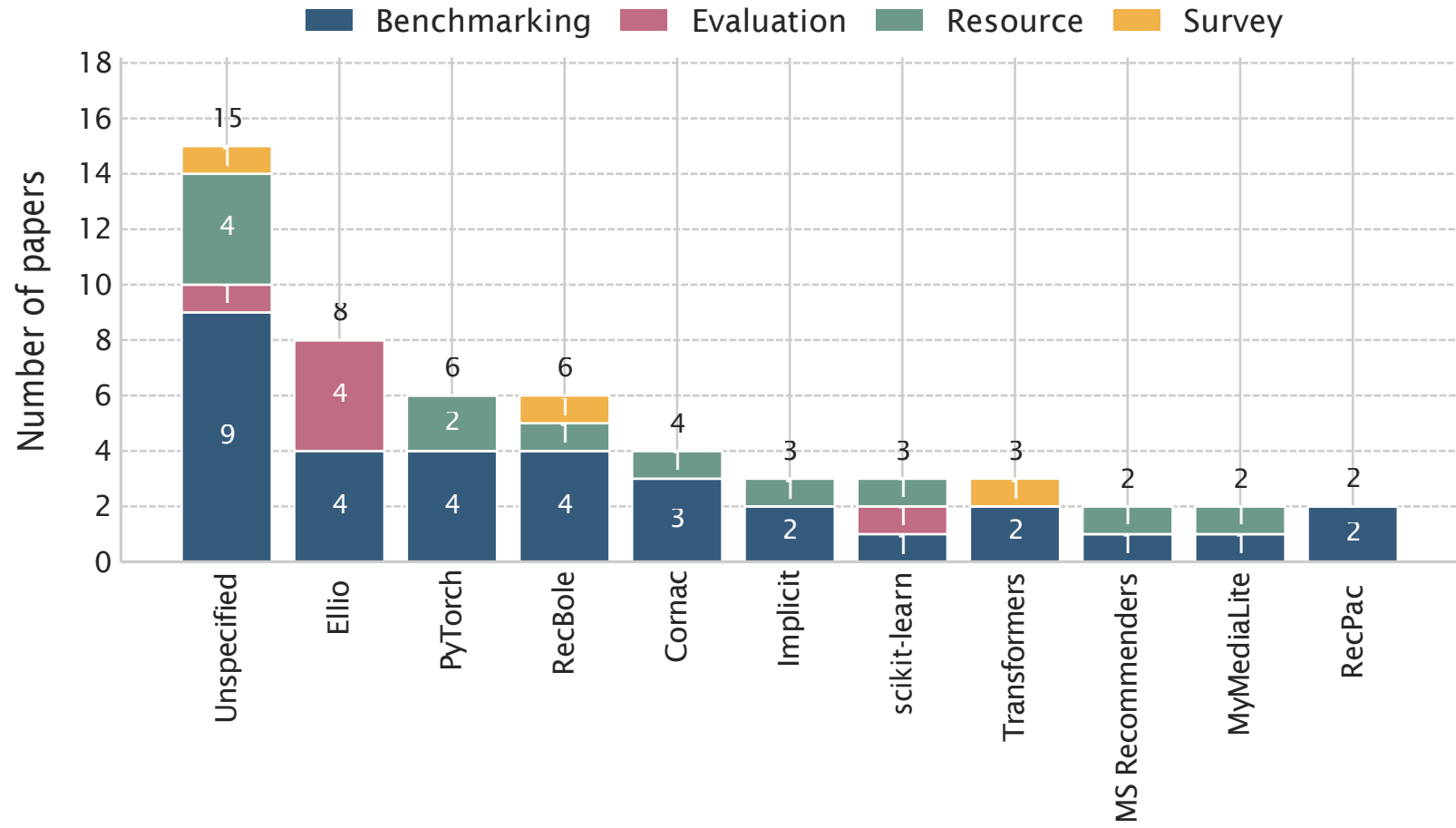

Fig. 4. Frequency of framework usage by paper type (as in Fig. 2). "Unspecified" indicates frameworks were not explicitly named. Only frameworks used in at least two papers shown.

Overall, framework usage reflects a partial standardization of experimentation practices: while many studies rely on shared tools, reporting practices remain inconsistent, limiting the ability to reproduce results across different implementations and environments.

### 3.3 Datasets

Datasets are foundational to empirical research in recommender systems, and their characteristics—such as domain, sparsity, and user–item interaction patterns—can significantly influence the outcome and generalizability of experiments. Our analysis shows that reproducibility papers rely heavily on a small number of dataset families, particularly MovieLens, Amazon, and LastFM (Fig. 5).

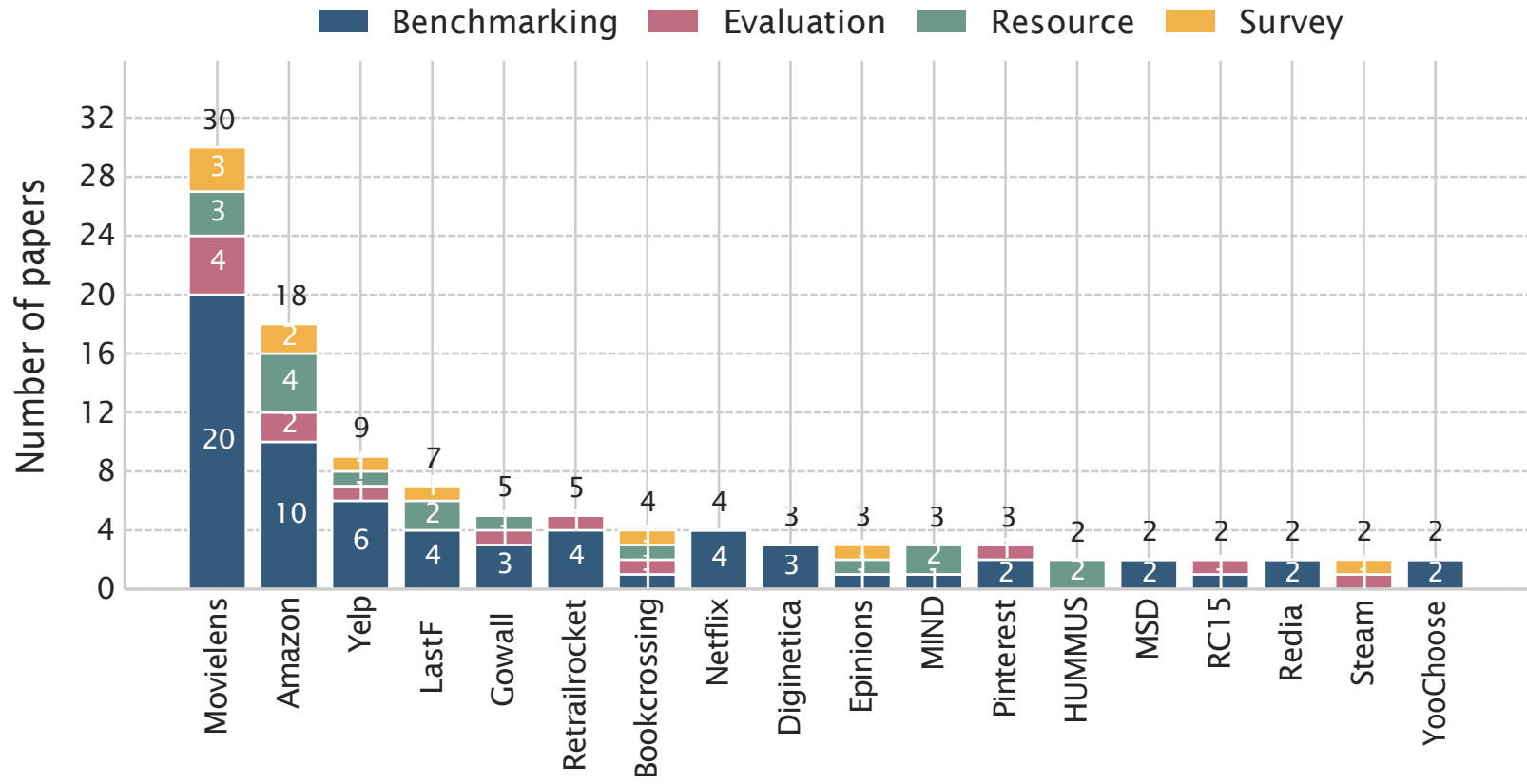

Fig. 5. Dataset usage by paper type (as in Fig. 2). Dataset families (e.g., Amazon, LastFM) are aggregated across variations. Only datasets used in at least two papers shown.

For readability, Fig. 5 aggregates multiple variants of popular dataset families into single categories. For example, the *Amazon* category includes datasets such as *Amazon Books*, *Amazon Electronics*, and *Amazon Games*; *MovieLens* encompasses *MovieLens 100K*, *1M*, and *10M*; and *LastFM* groups datasets such as *LastFM-1M* and *LastFM-2B*. While a more detailed analysis on the specific variants and sizes of these datasets would be interesting, since some papers simply state they use *MovieLens* (or *LastFM* or *Amazon*) with no more information, such analysis would not be possible. These datasets are frequently used in combination, as shown in Table 6, where common co-occurrences include (Amazon, MovieLens) and (MovieLens, LastFM). This reinforces their role as *de facto* benchmarks within the community. While this concentration facilitates comparison across studies, it also limits the diversity of experimental settings. Resource and benchmarking papers, in particular, rely heavily on these standard datasets, whereas evaluation and survey papers occasionally explore more specialized or domain-specific alternatives.

This pattern closely mirrors observations from prior studies of recommender systems research, where a small number of datasets dominate empirical evaluation [7]. As a result, reproducibility efforts often reinforce existing dataset choices rather than expanding the range of evaluated scenarios.

Furthermore, the prevalence of a small number of datasets raises questions about ecological validity and generalizability. While standardized datasets enable controlled comparisons, they may not fully capture the diversity of real-world recommendation contexts.

In this sense, and in line with **RQ3**, reproducibility studies tend to reinforce existing dataset choices rather than expanding the diversity of evaluation settings.

### 3.4 Algorithms

Algorithms constitute the core of reproducibility studies. Across the reviewed papers, we identified 186 distinct algorithmic instances, with a relatively small subset appearing frequently.

Despite this apparent diversity, algorithm usage is highly concentrated, with only a limited number of methods reused across multiple studies. Fig. 6 shows the most frequently used algorithms (appearing in at least two papers).

Among these, six algorithms stand out due to their frequent use: *MF*, *BPR*, *MostPop*, *GRU*-based models, *LightGCN*, and *NeuMF*. These methods collectively represent different generations and paradigms of recommender systems.

Table 6. Combinations of datasets frequently co-occurring in surveyed papers (with at least 3 occurrences).

| Dataset combinations | # Papers |
|---|---|
| (Amazon, Movielens), | 10 |
| (LastFM, Movielens), (Amazon, Gowalla), (Amazon, Yelp), (Movielens, Yelp), | 4 |
| (Amazon, Bookcrossing), (Amazon, Epinions), (Bookcrossing, Movielens), (Epinions, Movielens), (Amazon, Epinions, Movielens), (Movielens, Pinterest), (Gowalla, Yelp), (Amazon, Gowalla, Yelp), (Amazon, Netflix), (Movielens, Netflix), | 3 |

Table 7. Combinations of algorithms frequently co-occurring in surveyed papers (with at least 3 occurrences).

| Algorithm combinations | # Papers |
|---|---|
| (BPR, MostPop), | 9 |
| (ItemKNN, MostPop), (MostPop, NeuMF), (ALS, MostPop), (GRU, SASRec), (BPR, LightGCN), | 6 |
| (BPR, ItemKNN), (BPR, ItemKNN, MostPop), (EASE, MostPop), (MostPop, Mult-VAE), (LightGCN, MostPop), | 5 |
| (MostPop, Slim), (NeuMF, Slim), (MostPop, NeuMF, Slim), (ALS, NeuMF), (MostPop, RP3Beta), (LightGCN, NGCF), (LightGCN, SGL), (NGCF, SGL), (LightGCN, NGCF, SGL), (MostPop, Random), (LightGCN, NeuMF), (BPR, LightGCN, MostPop), (gpt, llama), | 4 |
| (BPR, NeuMF), (ItemKNN, NeuMF), (ALS, Slim), (MF, MostPop), (MF, NeuMF), (ALS, MostPop, NeuMF), (ALS, MostPop, Slim), (ALS, NeuMF, Slim), (MF, MostPop, NeuMF), (ALS, MostPop, NeuMF, Slim), (Bert4REC, SASRec), (EASE, NeuMF), (EASE, RP3Beta), (EASE, MostPop, RP3Beta), (MostPop, Mult-DAE), (Mult-DAE, Mult-VAE), (Mult-VAE, NeuMF), (MostPop, Mult-DAE, Mult-VAE), (MostPop, Mult-VAE, NeuMF), (DGCF, LightGCN), (ItemKNN, UserKNN), (MostPop, NGCF), (MostPop, UserKNN), (ItemKNN, MostPop, UserKNN), (NGCF, NeuMF), (ALS, Random), (BPR, Random), (ItemKNN, Random), (ALS, MostPop, Random), (BPR, ItemKNN, Random), (BPR, MostPop, Random), (ItemKNN, MostPop, Random), (BPR, ItemKNN, MostPop, Random), (CASER, GRU), (CASER, SASRec), (CASER, GRU, SASRec), (BPR, EASE), (BPR, NGCF), (BPR, SGL), (BPR, LightGCN, NGCF), (BPR, LightGCN, SGL), (BPR, NGCF, SGL), (BPR, LightGCN, NGCF, SGL), (BPR, Mult-VAE), (BPR, MostPop, Mult-VAE), | 3 |

*MostPop* serves as a simple but strong popularity-based baseline, often included as a lower bound for performance. *MF* encompasses various matrix factorization implementations (with or without user/item biases, exploiting implicit feedback or not, etc.) that learn user and item factors based on available data [30, 36]. *BPR*, although originally an optimization criterion [55], is commonly used as shorthand for matrix factorization models trained with pairwise ranking, and remains a widely adopted classical baseline. *GRU*-based models [29] represent early neural approaches to sequential recommendation, capturing temporal dependencies in user behavior. *NeuMF* exemplifies neural collaborative filtering methods that combine latent factor models with deep learning architectures [25], while *LightGCN* reflects more recent graph-based approaches [24] that simplify graph neural networks while maintaining strong performance.

The frequent co-occurrence of these methods (Table 7) reveals recurring evaluation setups centered around standard baselines and widely used models. These patterns suggest that reproducibility studies tend to compare a stable set of representative algorithms spanning baseline, classical, and neural approaches. Rather than exploring a broad algorithmic space, many studies focus on validating results within this established set, which effectively forms a de facto evaluation backbone in the literature.

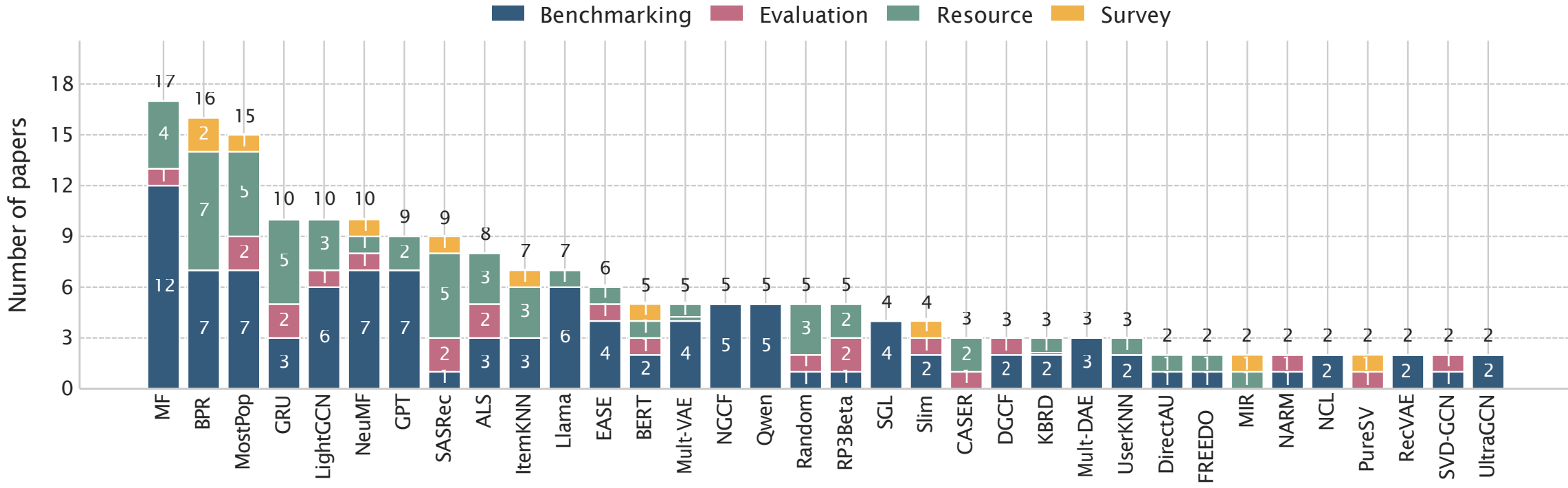

Fig. 6. Most frequently used recommendation algorithms, broken down by contribution type (as in Fig. 2). Only algorithms used in at least two papers shown.

To support aggregation, we grouped variations of the same model under a single label (e.g., different implementations of *BPR* or variants of *ALS* were unified under a shared identifier). This enables a clearer view of algorithmic reuse across papers, though it obscures fine-grained differences in implementation or optimization. We refer to our supplemental material for a fine-grained overview of specific algorithm use without aggregation.

Another important aspect of algorithm reproducibility is the transparency of configuration and training procedures. While some papers report hyperparameter settings explicitly or describe their tuning strategy, others rely on framework defaults or omit such details entirely. This inconsistency makes it difficult to determine whether observed performance differences stem from model implementation, parameterization, or dataset-specific effects.

Overall, the analysis highlights that reproducibility work covers a wide range of algorithms (Table 4), but the subset used in practice is relatively narrow (Fig. 6), with only six methods appearing ten or more times.

This concentration of algorithm choices further contributes to a standardized evaluation setup, reinforcing the methodological patterns discussed in RQ3.

### 3.5 Evaluation Practices

Evaluation strategies play an important role in reproducibility by determining how algorithmic performance is assessed and compared. We examined the evaluation practices adopted across reproducibility papers, focusing on two main aspects: the metrics used to report performance and the data splitting strategies employed for training and testing.

*Metrics.* Figure 7 presents an overview of the most frequently used evaluation metrics. Consistent with broader trends in recommender systems research, metrics such as *nDCG* and *Recall* are the most commonly reported, often in various cutoff versions (e.g., nDCG@10, Recall@20). Other metrics such as *Precision* and *Hit Rate* (HR) were also common, while beyond-accuracy metrics, such as diversity or fairness, appear only sporadically, and efficiency-related measures (e.g., prediction time) are included in a small number of papers. Table 8 shows that combinations of metrics are dominated by ranking and relevance measures, with limited integration of other evaluation dimensions.

While using similar metrics helps with comparing results across studies, it also highlights a tendency to focus on top-$k$ ranking accuracy, with relatively limited attention given to diversity, novelty, fairness, or robustness metrics, following similar observations made in, e.g., research on evaluation of recommender systems [7]. One notable observation is that error-based metrics, e.g., root mean squared error (RMSE) and mean absolute error (MAE), are not commonly used in the the sampled papers, indicating that error-based methods are not relevant for reproducibility purposes. This highlights the increasing disuse of error-based metrics in the recommender systems community, and the consideration that papers in this track aim to follow *evaluation best practices*, since in many cases the goal is to *improve* or *complete* the evaluation comparisons presented in previous works, especially in papers categorized as Benchmarking or Evaluation (see Table 2 and Table 5).

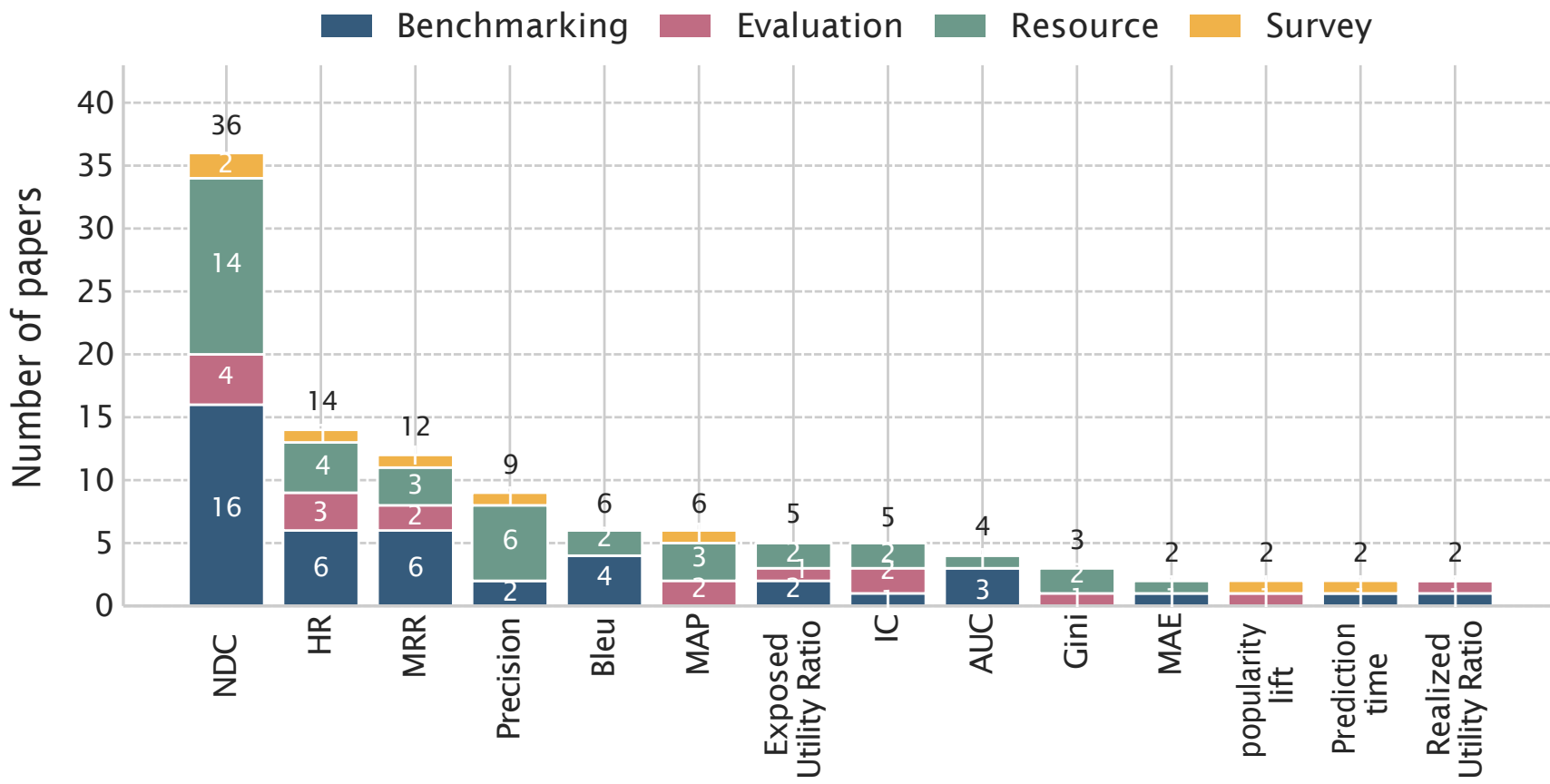

Fig. 7. Frequency of evaluation metrics across papers and types (as in Fig. 2). nDCG and Recall include multiple cutoff variants (e.g., nDCG@10). Only metrics used in at least two papers shown.

Table 8. Combinations of metrics frequently co-occurring in surveyed papers (with at least 3 occurrences).

| Metric combinations | # Papers |
|---|---|
| (NDCG, Recall), | 19 |
| (HR, NDCG), | 11 |
| (NDCG, Precision), (Precision, Recall), (NDCG, Precision, Recall), | 9 |
| (MRR, NDCG), | 7 |
| (HR, MRR), (MAP, NDCG), | 6 |
| (Bleu, Rouge), | 5 |
| (HR, MRR, NDCG), (F1, NDCG), (IC, NDCG), | 4 |
| (HR, MAP), (MAP, MRR), (MAP, Precision), (MAP, Recall), (MRR, Recall), (HR, MAP, NDCG), (MAP, MRR, NDCG), (MAP, NDCG, Precision), (MAP, NDCG, Recall), (MAP, Precision, Recall), (MAP, NDCG, Precision, Recall), (Gini, NDCG), (HR, IC), (AUC, NDCG), (Bleu, NDCG), | 3 |

*Data Splitting Strategies.* In addition to the choice of metrics, the way in which data is split for training and evaluation has significant implications for reproducibility [59]. As shown in Fig. 8, random splits remains the most popularly used method (11 out of 50). Temporal splits—whether sequential, global, or last-item-based—appear less frequently, despite being more realistic for time-aware recommendation scenarios. The preference for random splits may reflect a trade-off between ease of implementation and experimental fidelity, but it also raises concerns about ecological validity and over-optimistic performance estimates. Perhaps most notable is the lack of reporting of split type in the majority of papers (29).

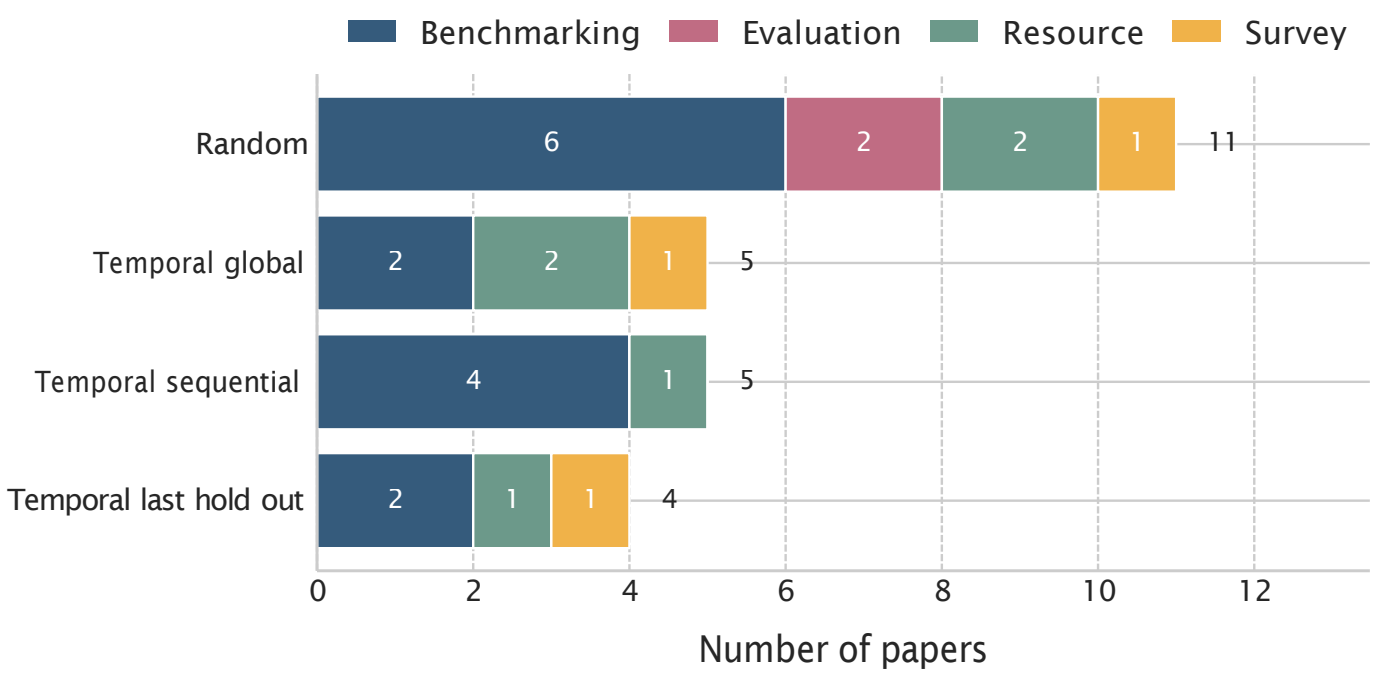

Fig. 8. Distribution of data splitting strategies by paper type (as in Fig. 2).

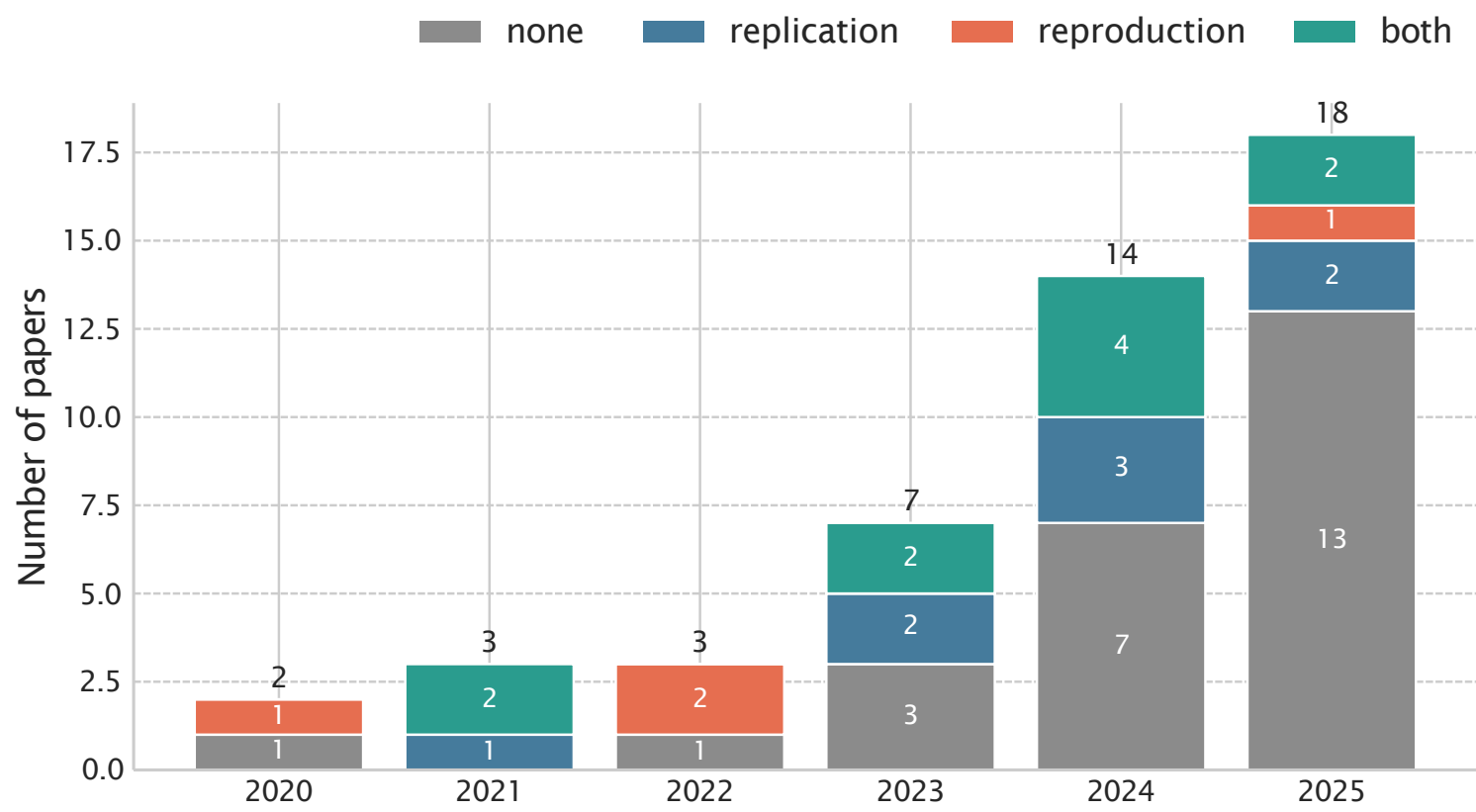

Fig. 9. Distribution of reproduction types by year (2020–2025) in the RecSys Reproducibility Track.

Taken together, these findings suggest that, addressing **RQ3**, reproducibility papers largely mirror established methodological patterns in recommender systems research rather than diverging from them.

### 3.6 Measuring reproducibility efforts

As discussed in Section 3.1, across 2020–2025, the RecSys Reproducibility Track exhibits both a substantial increase in submission volume and a shift in the nature of reproducibility contributions. Fig. 9 shows that in the early years (2020–2022), the track was dominated by reproduction and replication studies, with only a small number of papers falling outside these categories.

From 2023 onward, this distribution changes markedly. The number of papers not centered on reproduction or replication (denoted as *none*) increases steadily, reaching 7 papers in 2024 and 13 in 2025, while the number of reproduction and replication papers remains relatively stable. This shift coincides with the expansion of the track's scope (Section 2.1), which introduced resource, methodology, and reflective contributions. As a result, reproduction and replication studies become a minority, and the track evolves from a narrowly focused reproducibility venue into a broader forum for methodological and infrastructural research.

Beyond this quantitative shift, our analysis reveals important nuances in how reproducibility is operationalized. Several papers (e.g., [3, 4]) do not aim to fully reproduce prior work in its entirety—including all datasets, baselines, and configurations—but instead reimplement selected models, typically the main proposed method. Once this model is considered successfully reproduced, additional experiments are often conducted under modified conditions, effectively transitioning from reproduction to replication.

Table 9. Summary of reproducibility efforts in terms of numbers of removals (-) or additions (+) for algorithms, datasets, and metrics between papers from the Reproducibility track and the original paper being reproduced/replicated.

| Paper | Original Venue | Year | Reproducibility Paper | Year | Year gap | Summary of differences Algorithms | Datasets | Metrics |
|---|---|---|---|---|---|---|---|---|
| [25] | WWW | 2017 | [56] | 2020 | 3 | -2, +1 | -0, +0 | -0, +1 |
| [37] | KDD | 2020 | [18] | 2021 | 1 | -2, +4 | -0, +4 | -5, +1 |
| [15] | EMNLP | 2019 | [45] | 2021 | 2 | -2, +2 | -0, +0 | -4, +1 |
| [56] | RecSys | 2020 | [1] | 2021 | 1 | -0, +3 | -0, +0 | -1, +15 |
| [68] | CIKM | 2019 | [51] | 2022 | 3 | -6, +2 | -0, +0 | -1, +2 |
| [30] | ICDM | 2008 | [57] | 2022 | 14 | -2, +5 | -1, +4 | -1, +3 |
| [54] | SIGIR | 2020 | [39] | 2022 | 2 | -7, +4 | -0, +0 | -0, +0 |
| [23] | AAAI | 2016 | [42] | 2023 | 7 | -5, +1 | -3, +0 | -0, +0 |
| [67] | WSDM | 2022 | [50] | 2023 | 1 | -0, +0 | -0, +0 | -0, +3 |
| [29] | ICLR | 2016 | [28] | 2023 | 7 | -4, +5 | -2, +5 | -0, +0 |
| [40] | ISMIR | 2022 | [47] | 2024 | 2 | -0, +0 | -0, +1 | -1, +2 |
| [12] | CIKM | 2023 | [13] | 2024 | 1 | -8, +13 | -1, +4 | -0, +0 |
| [14] | AAAI | 2019 | [16] | 2024 | 5 | -4, +0 | -0, +0 | -0, +0 |
| [20] | RecSys | 2022 | [43] | 2024 | 2 | -16, +3 | -1, +0 | -0, +0 |
| [55] | UAI | 2009 | [48] | 2024 | 15 | -4, +10 | -1, +3 | -0, +2 |
| [77] | WSDM | 2024 | [19] | 2025 | 1 | -2, +3 | -1, +1 | -0, +9 |
| [72] | ECIR | 2025 | [71] | 2025 | 0 | -0, +0 | -1, +2 | -2, +1 |

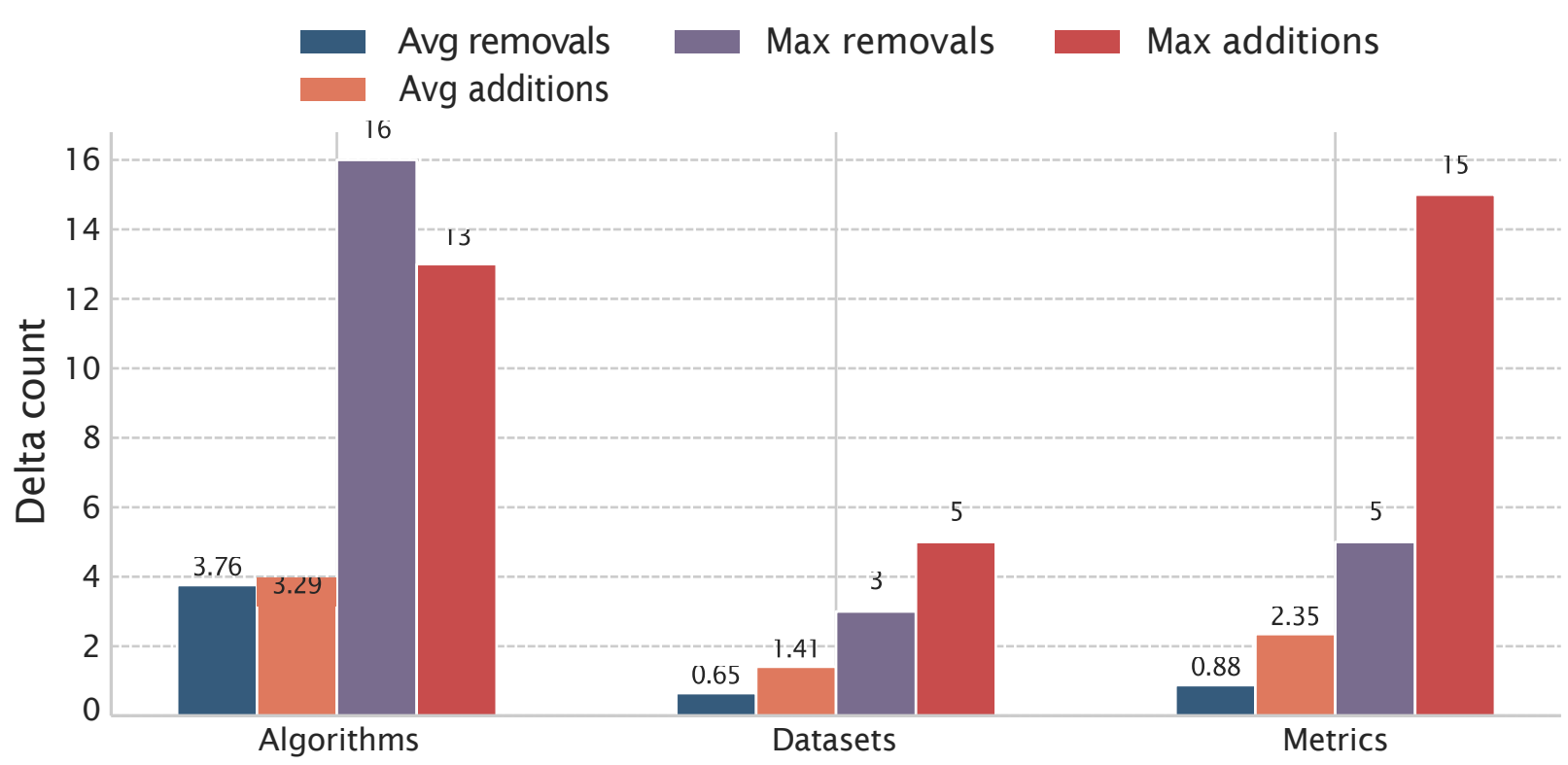

Fig. 10. Statistics about reproducibility efforts in terms of number of removals or additions for 3 dimensions: algorithms, datasets, metrics.

To capture this distinction, we identify in Table 3 the subset of papers that reproduce or replicate a single prior work. We exclude studies that benchmark multiple papers without fully reproducing their experimental setups, as these do not align with a strict definition of reproduction. For this subset, Table 9 and Fig. 10 quantify how reproducibility studies differ from the original work in terms of datasets, algorithms, and evaluation metrics. Three main patterns emerge.

First, algorithms are the most frequently modified component. Many reproducibility studies introduce additional baselines or alternative models, expanding the comparison space beyond the original work. Second, evaluation metrics are often added, typically to complement or strengthen the original evaluation protocol. This suggests that reproducibility efforts frequently aim to provide a more comprehensive assessment rather than a strict replication. Third, datasets tend to remain relatively stable. In many cases, the same datasets are reused, indicating that reproducibility studies prioritize comparability with the original work over exploring new data contexts.

These observations directly address **RQ4**, showing that reproducibility efforts most often involve extending and refining prior experiments rather than strictly replicating them.

We also observe that reproducibility studies are typically conducted shortly after the original publication. With only a few exceptions (e.g., gaps of 7, 14, or 15 years), most reproduced works are recent, suggesting that authors target papers that are both influential and still actively discussed in the community.

Taken together, these findings indicate that reproducibility in practice often involves extending and refining prior experiments rather than strictly replicating them. While this approach increases the breadth of evaluation and improves transparency, it also highlights a gap between the conceptual definition of reproducibility and its implementation in the RecSys community.

### 3.7 Summary

Across all dimensions, reproducibility papers exhibit a consistent methodological profile. Addressing **RQ1**, the RecSys Reproducibility Track has evolved from a narrowly focused venue for reproduction and replication studies into a broader forum encompassing benchmarking, resources, and methodological contributions.

In terms of methodological practice (**RQ2**), reproducibility papers rely on a limited set of frameworks, datasets, algorithms, and evaluation metrics, and frequently reuse established experimental configurations. While this consistency supports comparability across studies, it also reflects a lack of methodological diversity.

When interpreted in light of prior studies of recommender systems research [7, 58, 78], these findings indicate, addressing **RQ3**, that reproducibility efforts largely mirror existing evaluation practices rather than fundamentally challenging them.

Finally, addressing **RQ4**, reproducibility in practice often involves extending and refining prior experiments—frequently adding algorithms and metrics while keeping datasets stable—highlighting a gap between the conceptual definition of reproducibility and its implementation.

As a result, the primary contribution of reproducibility work lies in improving transparency, documentation, and empirical scrutiny, rather than expanding the methodological space of the field.

## 4 Discussion

Reproducibility is widely recognized in the RecSys community as a prerequisite for credible and cumulative research. At the same time, what counts as reproducibility in practice remains less clearly defined. Our analysis shows that the RecSys Reproducibility Track has evolved both in scale and in scope, moving from narrowly defined venue for reproduction studies toward broader forum that also includes methodological and infrastructural contributions. This shift has increased participation and diversified contributions, but it has also introduced a tension between the original intent of the track and its current role.

One observation that stands out is that the expansion of track in scope has not been followed by a corresponding diversification of methodological practice. While contributions have broadened, the underlying experimental setups remain stable. This creates a situation where the form of reproducibility research evolves, but its substance changes more slowly. We now discuss this tension across key dimensions in recommender systems implementation and experimentation.

The increasing use of **frameworks** such as Elliot, RecBole, and Cornac reflects a move toward more standardized experimentation. These frameworks make it easier to implement models, share pipelines, and lower the barrier to entry for reproducibility studies. In that sense, they represent clear progress. Differences in configuration, preprocessing, and default parameters can still lead to substantially different outcomes, even within the same framework. Moreover, a non-trivial number of papers do not clearly report the frameworks they rely on, which limits interpretability. Frameworks therefore improve reproducibility at the level of execution, but do not fully resolve the issue of comparability across different studies.

A similar pattern appears in **dataset** usage. A small number of datasets, most notably MovieLens, Amazon, and LastFM, dominate reproducibility studies and often function as reference points. This supports comparison across studies and makes it easier to relate results to prior work. At the same time, it constrains the range of experimental settings under which models are evaluated. Reproducibility studies frequently reuse the same datasets as the original work, which prioritizes comparability but limits the ability to assess generalizability. As a result, reproducibility efforts tend to reinforce existing dataset choices rather than extend evaluation to new domains

**Algorithm** selection follows the same logic. While a large number of models appear across papers, a relatively small subset is used repeatedly. Methods such as MF, BPR, MostPop, GRU-based models, NeuMF, and LightGCN appear consistently and span different generations of recommender systems research. This shared set of baselines and reference models provides structure and consistency, making results easier to interpret in relation to prior work. At the same time, it narrows the scope of experimentation. New or less common approaches are rarely included, and reproducibility studies tend to operate within an already established set of algorithms, reinforcing existing evaluation practices rather than expanding them.

Evaluation practices further reinforce this patterns. **Metrics** such as nDCG, Recall, and Precision dominate, while beyond-accuracy measures remain underrepresented even as they receive increasing attention in recommender systems research. Similarly, **data splitting** strategies remain dominated by standard approaches such as random splits, despite long-standing recommendations favoring temporal evaluation protocols for many recommendation tasks [35]. One possible explanation is that reproducibility studies often prioritize comparability with the original work over adopting improved evaluation methodologies. Since reproducing published results typically requires matching the original experimental setup, established practices, including suboptimal splitting strategies, are naturally carried forward. This may help explain why methodological recommendations have had limited impact on experimental practice, even within a venue explicitly focused on reproducibility. These choices mirror broader trends in recommender systems research. Reproducibility studies improve transparency in how evaluation is conducted, but they rarely question or revise established paradigms.

Perhaps the most important observation is how reproducibility is approached **in practice**. Most reproducibility studies do not strictly aim to exactly reproduce the full experimental setup of prior work. Instead, they typically reimplement the main core model and then extend the evaluation by introducing additional baselines, metrics, or experimental variations. Once the model is reproduced, the study often shifts toward comparison rather than strict replication. This reflects a pragmatic approach in which reproduction serves as a starting point for further analysis. While this increases the breadth of evaluation and provides more comprehensive comparisons, it also introduces a gap between the conceptual definition of reproducibility and its implementation. If reproducibility is defined as exact repetition, many studies only partially meet this criterion. If it is defined more broadly as revisiting and extending prior work, then current practice aligns with that interpretation.

Taken together, these results suggest that reproducibility work has improved transparency, documentation, and empirical scrutiny. It has made it easier to understand, verify, and build upon prior work. However, its impact on methodological diversity is limited. The same datasets, algorithms, and evaluation practices continue to dominate, and reproducibility studies tend to consolidate rather than challenge these patterns. This suggests that reproducibility, as currently practiced, supports consolidation of existing methods more than exploration of new ones.

This points to a potential next step for the community. Reproducibility could play a more active role not only in confirming existing results, but also in systematically evaluate them under different conditions. This includes evaluating models on less commonly used datasets, incorporating a broader set of metrics, and exploring alternative experimental designs. Such efforts would move reproducibility beyond verification toward a more critical and exploratory role in shaping research practices.

## 5 Conclusions and Future Work

This paper presented a structured analysis of the RecSys Reproducibility Track from 2020 to 2025, covering 51 accepted papers. We examined how reproducibility is operationalized in practice by analyzing contribution types, datasets, algorithms, frameworks, and evaluation protocols, as well as how studies reproduce or extend prior work.

We observed three main conclusions. First, the scope of the track has expanded substantially. While early editions focused on reproduction and replication, more recent years include a broader range of contributions such as benchmarking, resources, and methodological work. This expansion has increased participation and diversity, but has also shifted the center of gravity away from strict reproduction studies.

Second, reproducibility papers exhibit a consistent methodological profile. A small set of datasets, algorithms, frameworks, and evaluation practices dominate across studies. This supports comparability, but also indicates that reproducibility work largely operates within established experimental conventions rather than expanding them.

Third, reproducibility in practice often takes the form of extension rather than strict replication. Studies typically reimplement core models and then introduce additional algorithms, metrics, or experimental variations. While this improves empirical coverage and transparency, it highlights a gap between the conceptual definition of reproducibility and its implementation.

Taken together, these results suggest that reproducibility efforts in recommender systems have been effective in improving transparency and documentation, but have had a more limited impact on methodological diversity. Reproducibility has become a mechanism for consolidating existing practices rather than systematically challenging them.

Future work should build on this foundation by broadening the scope of reproducibility. This includes evaluating models across more diverse datasets, incorporating beyond-accuracy metrics, and exploring alternative experimental designs. In addition, clearer guidelines on what constitutes reproduction versus replication would help align expectations and practices within the community.

### 5.1 Limitations

This study has some limitations. First, it focuses exclusively on the RecSys Reproducibility Track, which may not fully represent reproducibility practices in recommender systems research more broadly. Second, our analysis relies on reported experimental details, which may omit implementation-specific factors that affect reproducibility. Third, while our coding schema captures key methodological aspects, it necessarily abstracts away from fine-grained differences in model implementations and experimental configurations.

### 5.2 Future Directions

Future research could extend this work in several directions. A longitudinal analysis beyond the RecSys track could provide a broader view of reproducibility practices across venues. Further work could also investigate the relationship between reproducibility and evaluation quality, particularly in terms of beyond-accuracy metrics and real-world deployment scenarios. Considering the framework space is relatively standard, even though no framework clearly dominates the others (and more implementations are made public every year), it is necessary to conduct an experimental comparison considering the most popular frameworks, datasets, or metrics to assess if the reproducibility efforts invested during these years have had an effect on the actual reproducibility or comparability of results. Finally, developing standardized evaluation pipelines and reporting practices may help bridge the gap between reproducibility and comparability across studies.

## Acknowledgments

Work supported by grant PID2022-139131NB-I00 funded by MCIN/AEI/10.13039/501100011033 and by "ERDF A way of making Europe."